\documentclass[amsmath,amssymb]{revtex4-2}

\usepackage{graphicx}
\usepackage{dcolumn}
\usepackage{bm}

\begin{document}


\title{Microscopic insights of magnetism in Sm$ _{2} $NiMnO$ _{6} $ double perovskite}

\author{Supriyo Majumder$^{a}$}
\author{Malvika Tripathi$^{a,b}$}
\author{H. E. Fischer$^{c}$}
\author{D. O. de Souza$^{d}$}
\author{L. Olivi$^{d}$}
\author{A. K. Sinha$^{e,f}$}
\author{R. J. Choudhary$^{a}$}\email{ram@csr.res.in}
\author{and D. M. Phase$^{a}$}
\renewcommand{\andname}{\ignorespaces}
\affiliation{$^{a}$UGC DAE Consortium for Scientific Research, Indore 452001, India}
\affiliation{$^{b}$Niels Bohr Institute, University of Copenhagen, 2100 Copenhagen, Denmark}
\affiliation{$^{c}$Institut Laue-Langevin, 38042 Grenoble Cedex, France}
\affiliation{$^{d}$Elettra Sicrotrone Trieste S.C.p.A., SS 14-km 163.5, 34149 Basovizza, Italy}
\affiliation{$^{e}$HXAL, SUS, Raja Ramanna Centre for Advanced Technology, Indore 452013, India}
\affiliation{$^{f}$Department of Physics, School of Engineering UPES, Dehradun 248007, India}

\begin{abstract}
The functional characteristics of double perovskites with unique ferromagnetic-insulator ground state have been controversial due to the unavoidable presence of anti-site disorders (ASDs). Here, we aim to investigate the origin of magnetic ordering on local and global scales in Sm$_{2}$NiMnO$_{6}$ (SNMO) double perovskite system. Different calcination routes are exploited to generate different cation arrangements in SNMO and the corresponding magnetic configurations are examined using the high energy (E $ \sim $0.3 eV) `hot neutrons', which has helped to overcome Sm absorption as well as to record total (Bragg's+diffuse) scattering profiles with high momentum transfer (Q$ _{max} \sim $24 $ \AA^{-1} $). We have observed that the Ni-Mn sublattice adopts long range collinear ferromagnetic $ F_{x}F_{z} $ structure with commensurate \textbf{k}=(0, 0, 0) propagation vector, below ordering temperature T $ \lesssim $ 160 K, irrespective of variable ASD concentrations. In addition, the signatures indicating the antiparallel polarization of Sm paramagnetic moments with respect to Ni-Mn network, are noticed in the vicinity of anomalous magnetic transitions at T $ \lesssim $ 35 K. The real space pair distribution function calculations have provided a direct visualization of ASDs by means of broadening in Ni/Mn-Mn/Ni linkage. Employing the Reverse Monte Carlo approach on diffuse magnetic scattering profiles, we have observed the negative spin-spin correlation function which suggests the Ni-Ni antiferromagnetic exchange interactions ranging up to first nearest neighbor distance. These results confirm that the existence of ASDs in cation ordered host matrix leads to competing ferromagnetic-antiferromagnetic phases in a broad temperature range, which quantitatively governs the temperature dependent bulk magnetic observables of SNMO system.
\end{abstract}

\maketitle

\section{INTRODUCTION}
The existence of disorders is an unavoidable reality in every crystal system. Structural disorders are usually suspected to influence the heat conductivity \cite{ABalandin2011}, charge transport \cite{RNoriega2013}, magnetic exchange interactions \cite{JBowles2013} and many other functional properties in materials. The double perovskite oxide family A$_{2}$B$'$B$''$O$_{6}$ (where, A: mainly alkaline earth or rare earth; B$'$, B$''$: transition metal) \cite{MTAnderson1993}, is one of the most dramatic example of how significantly the structural dislocations can affect the electronic and magnetic ground states of the system. In double perovskites, the presence of anti-site disorders (ASDs), defined as the interchange of different B$'$, B$''$ cations from their respective alternate site occupancies, can even stimulate a transformation from ferromagnetic (FM) - metal (M) to antiferromagnetic (AFM) - insulator (I) as a special case \cite{MGHernandez2001, DDSarma2001}. 

The A$_{2}$NiMnO$_{6}$ (ANMO) double perovskites are the promising candidates for new generation dissipationless quantum electronics owing to the unique FM-I ground state \cite{DChoudhury2012, NSRogadol2005}. Furthermore, the near room temperature magneto-dielectric, spin pumping, magneto-resistance, magneto-capacitance behaviors \cite{DChoudhury2012, NSRogadol2005, YShiomi2014} and multiferroic properties \cite{MAzuma2005, JSu2015} have aroused renewed fundamental scientific interest in the double perovskites with Ni and Mn as B-site cations. However, the nearly similar ionic radii of Ni and Mn intrinsically leads to ASDs in ANMO double perovskite systems. Noteworthily, the characteristic electronic and magnetic behaviors of ANMO systems vary significantly among different reports available in literature \cite{WZYang, FGheorghiu}. These discrepancies can be correlated with the fact that different growth kinetics can lead to distinct cationic arrangement, which could eventually modify the physical observables \cite{SMajumder2022}. As a consequence, the reliability and reproducibility related to any functional aspects of ANMO series is questionable which limits their feasibility in device applications. Moreover, there are several unsettled fundamental issuses regarding the observed physio-chemical properties of ANMO system. For instance, contradicting explanations are found about the charge state of transition metal ions participating in magnetic interactions. Goodenough \textit{et al.} \cite{JBGoodenough1961} explained the observed magnetic behavior considering Ni$ ^{3+} $-O-Mn$ ^{3+} $ superexchange interactions. Whereas, Blasse \textit{et al.} \cite{GBlassel1965} claimed that magnetic ordering is due to Ni$ ^{2+} $-O-Mn$ ^{4+} $ superexchange interaction. On the other hand, both theoretical and experimental evidences suggest that the magnetic structure in ANMO series changes considerably for different A-site ions \cite{SKumar2010, WYi2013} and can vary among collinear FM (A=La, Nd and Y) \cite{NSRogadol2005, SanchezBenitez2011, HNhalil2015}, canted ferrimagnetic (A=Tb, Ho, Er and Tm) \cite{MRetuerto2015} and incommensurate AFM structure (A=In) \cite{NTerada2015}. As a special case, A=Sm leads to a complicated temperature driven magnetic phase diagram due to the unique intermixing of Sm$^{3+}$ higher multiplet levels (J= 7/2, 9/2 etc.) with the ground state multiplet (J=5/2) \cite{KHJBuschow1974, HAdachi1999}. In Sm$_{2}$NiMnO$_{6}$ (SNMO) system, there are several discrepancies among different reports regarding the origin of anomalous downturn in magnetic susceptibility below T=35 K. The possible explanations of the aforementioned magnetic behavior in SNMO include: (i) magnetocrystalline anisotropy governed by spin-orbit coupled Sm and Ni-Mn network \cite{RJBooth2009}, (ii) long range ordering of Sm sublattice moments \cite{WZYang}, and (iii) emergence of reentrant spin glass phase \cite{PNLekshmi2013}. The presence of ASDs introduce additional perturbation in the magnetic exchange interaction pathways which causes further complexity in ground state. Despite several research endeavors, how local ASDs can reshape the physical properties in SNMO double perovskite system, is still an open question.

One of the main reasons behind these inconsistencies are that the microscopic magnetic structure of SNMO system, and its thermal evolution are still not available in literature. In this context, neutron diffraction is a powerful probe to study the elemental spin arrangements. On the other hand, diffuse neutron scattering could be very useful to get the insight of local spin-spin correlations arising due to ASDs. However, very high neutron absorbing $ ^{149} $Sm isotope (absorption $ \sim $ 42080b \cite{JLynn1990, VFSears1992}) present in natural Sm, creates hindrance to perform neutron scattering experiments. The huge neutron absorption of $ ^{149} $Sm isotope arises possibly due to low energy ($ \sim $0.098 eV) nuclear resonances \cite{JLynn1990, PDeBievre1993}. The magnitude of neutron absorption by $ ^{149} $Sm can be decreased significantly, if the energy of the incident neutron beam can be tuned to a value sufficiently higher than the nuclear resonance energy. Therefore, to overcome the high neutron absorption, we have utilized high-energy ($ \lambda \sim $0.5 $ \AA $) `hot neutrons' to record diffraction profiles at D4 disordered materials diffractometer instrument \cite{HFischer2002}, Institut Laue-Langevin (ILL). The high momentum transfer (Q$ _{max} \sim $24 $ \AA^{-1} $), very low  background contribution, high counting rate, and temperature stability of D4 instrument has allowed us to record total (Bragg's+diffuse) scattering patterns with good statistics. In this work, we have constructed the magnetic phase diagram of SNMO double perovskites having different degree of cation arrangements, and explored the role of ASDs in governing the local and global magnetic structures.

\begin{figure*}[t]
\centering
\includegraphics[angle=0,width=0.92\textwidth]{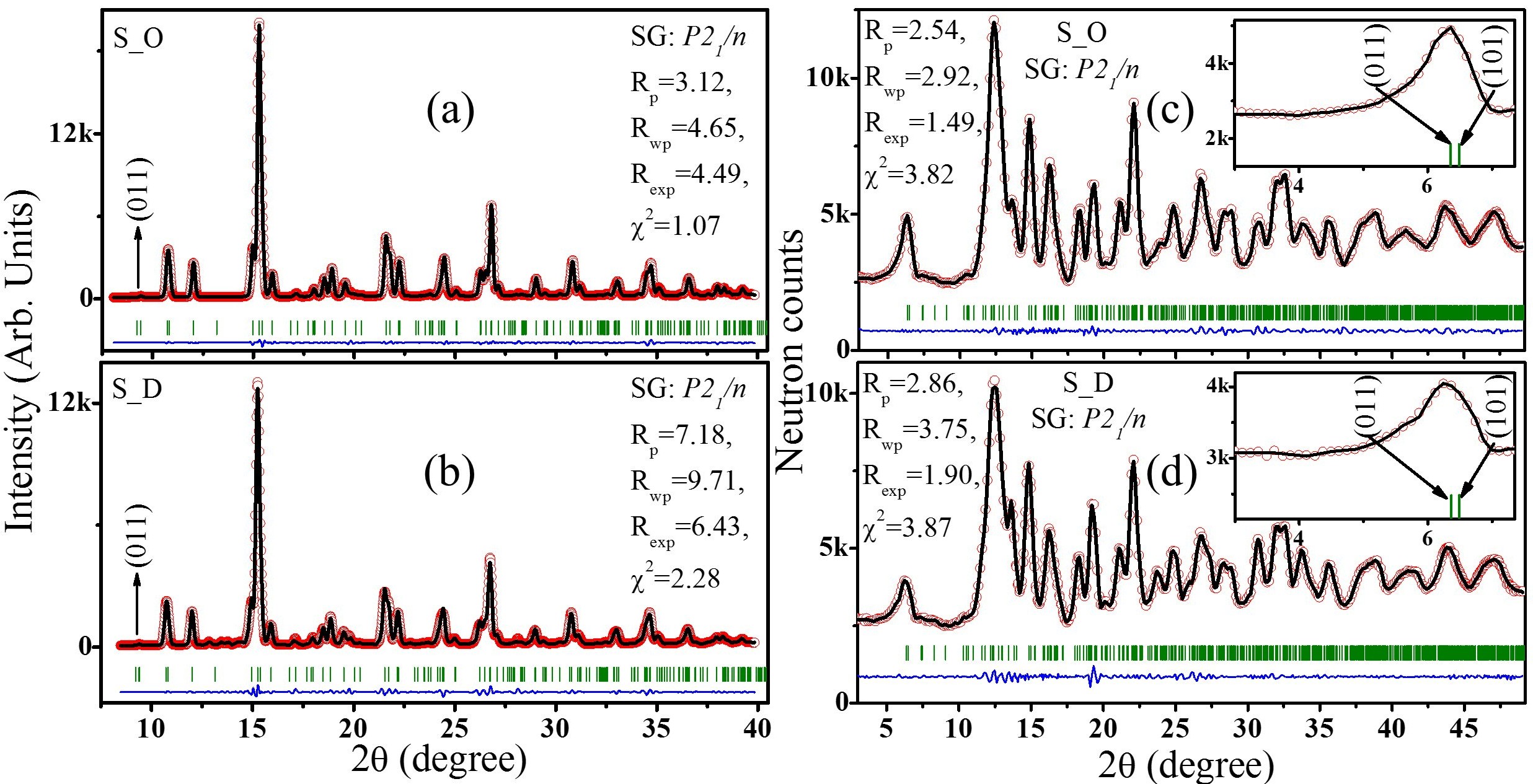}
\caption{Rietveld analysis of (a, b): PXRD and (c, d): NPD $2\theta$ scans measured at T=300 K, showing observed (red open circles), calculated (black solid line) and difference (blue solid line) pattern along with Bragg's positions (green vertical bars), for cation ordered S$\_$O and disordered S$\_$D SNMO samples. Insets of (c, d) show enlarged view of (011) superstructure reflection. The Rietveld generated patterns are simulated with monoclinic \textit{P2$_{1}$/n} (SG 14) structural model. The goodness of fitting indicators, for S$ \_ $O PXRD data are R$_{p}$=3.12, R$_{wp}$=4.65, R$_{exp}$=4.49, $\chi^2$=1.07; for S$ \_ $O NPD data are R$_{p}$=2.54, R$_{wp}$=2.92, R$_{exp}$=1.49, $\chi^2$=3.82; for S$ \_ $D PXRD data are R$_{p}$=7.18, R$_{wp}$=9.71, R$_{exp}$=6.43, $\chi^2$=2.28 and for S$ \_ $D NPD data are R$_{p}$=2.86, R$_{wp}$=3.75, R$_{exp}$=1.90, $\chi^2$=3.87.}\label{xrdnpdstr}
\end{figure*}

\section{EXPERIMENTAL DETAILS}
Polycrystalline SNMO bulk samples were synthesized by solid state reaction route. Different reaction conditions are utilized to achieve distinct ASD density. Stochiometric amount of precursors Sm$ _{2} $O$ _{3} $ (99.9\%), NiO (99.99\%) and MnO$ _{2} $ (99.9\%) were combined to form a homogeneous mixture which was further divided into two batches. The first batch was calcined at 1200$ ^{o} $C for 24h and slowly cooled at a rate $ \sim $1.6$ ^{o} $C/min (sample S$ \_ $O). The next batch was calcined at 1000$ ^{o} $C for 24h and thermally quenched at a rate of $ \sim $11.6$ ^{o} $C/min down to 650$ ^{o} $C and then cooled at a rate of $ \sim $3.3$ ^{o} $C/min (sample S$ \_ $D). In order to identify the room temperature phase composition and the crystal structure of SNMO, powder X-ray diffraction (PXRD) and neutron powder diffraction (NPD) $2\theta$ scans were recorded. PXRD studies were carried out using synchrotron radiation source at angle dispersive X-ray diffraction (ADXRD) beamline, BL-12, Indus-II, RRCAT, Indore, India. In the PXRD measurements, the calibrated wavelength using diffraction pattern of LaB$ _{6} $ standard sample was $ \lambda $=0.71949 $ \AA $. NPD data were collected \cite{MTripathi2019illdata} utilizing two-axis diffractometer D4 (Disordered Materials Diffractometer) \cite{HFischer2002} at ILL, Grenoble, France. Incident wavelength of neutron beam was calibrated by recording data for standard Ni sample and the estimated value was $ \lambda $=0.49585 $ \AA $. D4 provides neutron diffractogram measurements covering momentum-transfer Q range of $ \sim $0.3-23.7 $ \AA^{-1} $ ($ 2\theta $ range $ \sim $1.3-138.1$ ^{o} $) with relative Q resolution $ \sim $2.25$ \% $. Temperature dependent neutron diffraction studies were performed using cylindrical vanadium sample container and cryostat setup. Diffraction patterns for empty cylindrical vanadium container and cryostat assembly were collected at various temperature values. These allow us to accurately determine temperature dependent (if any) sample environment contributions which is necessary for pair distribution function (PDF) analysis. Using the CORRECT program \cite{MAHowe1996}, measured neutron diffraction intensity was corrected for background subtraction, sample attenuation, multiple scattering, inelastic (Placzek) scattering effects and normalized to absolute scale. Obtained total scattering intensity was used for PDF calculation. Crystal and magnetic structure refinement of PXRD and NPD data were performed implementing FULLPROF package \cite{JRodriguezCarvajal1993} wherein WINPLOTR was used for background simulation, BASIREPS was used for calculating the irreducible representations (IRs) and FPSTUDIO was used to visualize the obtained structure. A value of +5.0 fm was used for the coherent neutron scattering length of Sm at $ \lambda\sim $0.5 $ \AA $ as tabulated by Lynn and Seeger \cite{JLynn1990}, from whom we also obtained the Sm absorption cross-section for our neutron wavelength. SPINCORREL and SPINVERT programs \cite{JAMPaddison2013} based on Reverse Monte Carlo (RMC) approach were used for analyzing magnetic diffuse scattering data. Chemical valency of constituting elements was probed by X-ray absorption near edge spectroscopy (XANES) measurements performed at room temperature in transmission mode using synchrotron radiation source at the XAFS beamline 11.1R, Elettra-Sincrotrone, Trieste, Italy. Incident energy was calibrated by measuring absorption edges for reference Mn and Ni metal foils. The energy resolution $ \Delta E / E $ across the energy range used for XANES measurements was $ \sim $1$ \times $10$ ^{-4} $. Standard normalization process was performed on XANES data using ATHENA program \cite{BRavel2005}. Dc magnetization behavior were studied using MPMS 7-Tesla SQUID-VSM (Quantum Design Inc., USA) instrument having magnetic moment sensitivity $ \sim $10$ ^{-8} $ emu. To eliminate the trapped magnetic field of the superconducting coil inside the magnetometer, standard \textit{de}-\textit{Gaussing} protocol was followed. To erase the prior magnetic history (if any), the sample was heated well above its magnetic transition temperatures every time before recording the data.

\begin{table*}[]
\centering
 	\begin{tabular}{@{}ccccc@{}}
 		\hline
 		 & \multicolumn{2}{c}{S$\_$O} & \multicolumn{2}{c}{S$\_$D} \\
 		 & PXRD & NPD & PXRD & NPD \\
 		\hline
 		Lattice parameters ($\AA$) & & & & \\
 		a & 5.35647(11) & 5.35939(53) & 5.35563(27) & 5.35648(61) \\
 		b & 5.52127(11) & 5.52169(49) & 5.51294(26) & 5.51365(52) \\ 
 		c & 7.61932(17) & 7.61924(59) & 7.61326(39) & 7.61893(97) \\
 		$ \beta $ & 90.030(26) & 90.031(12) &  90.030(26) & 90.030(15) \\
 		Site occupancy $\&$ & & & & \\ 
		Fractional co-ordinates & & & & \\ 
 		Sm(\textit{4e}) & 1.0 & 1.001(5) & 1.0 & 0.999(6) \\
 		x & -0.00978(14)& -0.01088(80) & -0.01115(26) & -0.01107(122)  \\
 		y & 0.05580(7)& 0.05604(43) & 0.05604(14) & 0.05614(79) \\
 		z & 0.25053(31)& 0.25005(52) & 0.24933(57) & 0.25150(77) \\
 		Ni$|$Mn(\textit{2c}) & 0.5 & 0.481(1)$|$0.018(5) & 0.5 & 0.440(2)$|$0.060(6) \\
		x & 0.5 & 0.5 & 0.5 & 0.5 \\
 		y & 0.0 & 0.0 & 0.0 & 0.0 \\
 		z & 0.5 & 0.5 & 0.5 & 0.5 \\
 		Mn$|$Ni(\textit{2d}) & 0.5 & 0.480(3)$|$0.020(1) & 0.5 & 0.442(4)$|$0.059(1) \\
		x & 0.5 & 0.5 & 0.5 & 0.5 \\
 		y & 0.0 & 0.0 & 0.0 & 0.0 \\
 		z & 0.0 & 0.0 & 0.0 & 0.0 \\
 		O1(\textit{4e}) & 1.0 & 0.994(8) & 1.0 & 1.000(9) \\
 		x & 0.09535(103) & 0.09446(51) & 0.07667(203) &  0.07329(68) \\
 		y & 0.47529(88) & 0.47516(51) & 0.47414(224) & 0.47740(64)   \\
 		z & 0.22849(139) & 0.22949(36) & 0.23245(196) & 0.23383(44)  \\
 		O2(\textit{4e}) & 1.0 & 0.998(7) & 1.0 & 0.991(7) \\
 		x & 0.72503(200) & 0.72815(57) & 0.66045(525) & 0.67398(71)  \\
 		y & 0.26758(171) & 0.26824(40) & 0.24234(328) & 0.25174(73)  \\
 		z & 0.03219(211) & 0.03612(43) & 0.00535(178) & 0.01260(54)  \\
 		O3(\textit{4e}) & 1.0 & 0.994(9) & 1.0 & 1.004(8) \\
 		x & 0.69691(175) & 0.70041(53) & 0.69099(172) & 0.69126(61) \\
 		y & 0.29364(124) & 0.29330(55) & 0.32438(199) & 0.32333(63)  \\
 		z & 0.45827(147) & 0.45825(56) & 0.43036(159) & 0.43290(44)  \\
 		Isotropic thermal factors ($\AA^2$) & & & & \\
 		$\it{B}_{Sm}$ & 0.072(9) & 0.100(31) & 0.068(17) & 0.121(45) \\
 		$\it{B}_{Ni}$ & 0.049(39) & 0.089(17) & 0.057(28) & 0.091(27) \\
 		$\it{B}_{Mn}$ & 0.046(39) & 0.123(51) & 0.054(24) & 0.110(79) \\
 		Statistical parameters & & & & \\
 		R$_{p}$ & 3.12 & 2.54 & 7.18 & 2.86 \\
 		R$_{wp}$ & 4.65 & 2.92 & 9.71 & 3.75 \\
 		R$_{exp}$ & 4.49 & 1.49 & 6.43 & 1.90 \\
 		$\chi^2$ & 1.07 & 3.82 & 2.28 & 3.87 \\
 		\hline\hline
 	\end{tabular}
 	\caption{Structure parameters and reliability indicators obtained from Rietveld refinement of PXRD and NPD data recorded at T=300 K for SNMO, S$\_$O: ordered and S$\_$D disordered samples. Number(s) in the parentheses represent the error bars on the last digit(s). Here, the site occupancies and fractional coordinates which do not have error bars are fixed to crystallographic values.}
 	\label{pxrdnpd300krefinetable}
\end{table*}

\begin{figure*}[t]
\centering
\includegraphics[angle=0,width=0.8\textwidth]{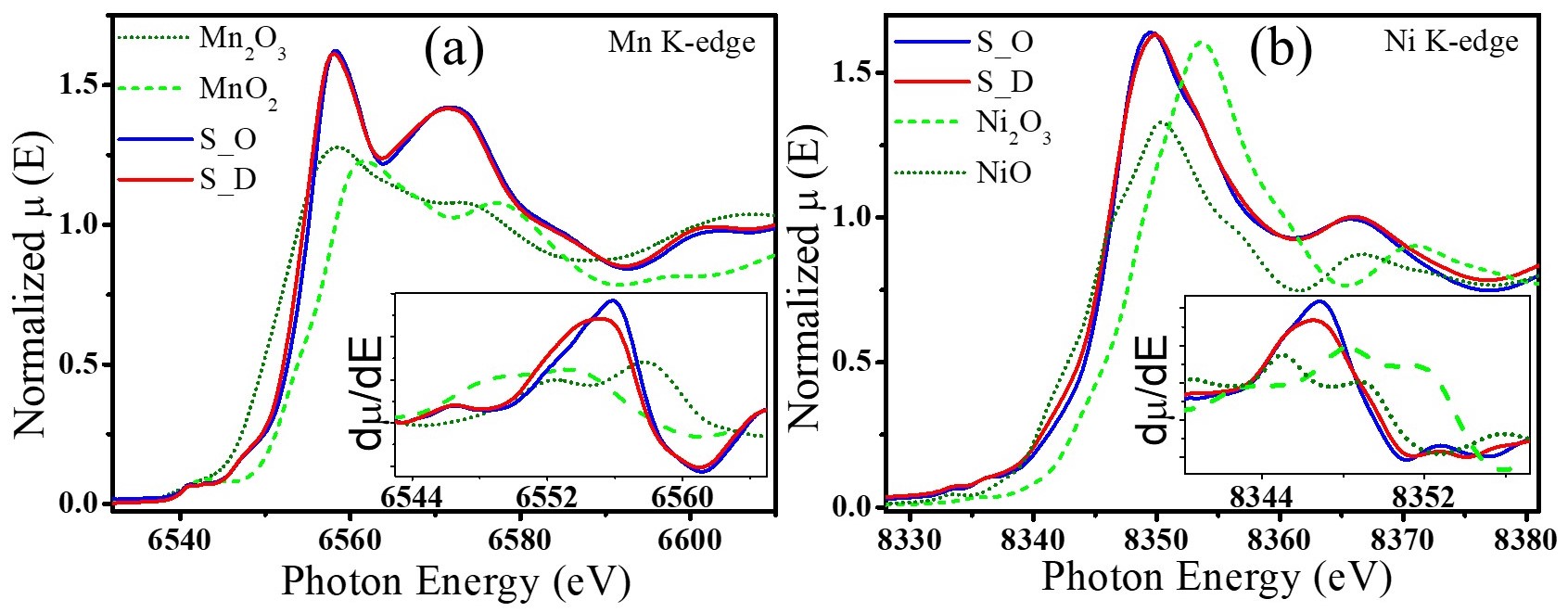}
\caption{(a): Mn and (b): Ni \textit{K} edge XANES spectra measured at T=300 K for SNMO samples along with standard references. Insets show first derivative of normalized absorption with respect to incident photon energies.}\label{xanes}
\end{figure*}

\section{RESULTS AND DISCUSSION}
Room temperature powder X-ray diffractograms of SNMO samples are presented in Figs. \ref{xrdnpdstr}(a, b). These patterns can be matched with simulated Rietveld profiles generated from both monoclinic \textit{P2$_{1}$/n} (SG 14) and orthorhombic \textit{Pbnm} (SG 62) structures. This is due to fact that the unit cell $ \surd 2a_{p} \times \surd 2a_{p} \times 2a_{p} $, where $a_{p}$ stands for pseudo-cubic, is subjected to both \textit{P2$_{1}$/n} and \textit{Pbnm} space groups depending on B-site ordered and disordered phases, respectively. In \textit{P2$_{1}$/n} symmetry there are two independent Wyckoff positions for B$'$ and B$''$ ions, \textit{2c} and \textit{2d}, respectively. Whereas, in the \textit{Pbnm} symmetry, both B$'$ and B$''$ ions randomly occupy identical \textit{4b} site. Although, \textit{P2$_{1}$/n} is maximal non-isomorphic subgroup of \textit{Pbnm}, the presence of $ (0kl) $: $ k=2n+1 $ superstructure reflections which is an indication for B$'$, B$''$ cation ordering at alternate octahedral center sites, can be used to distinguish space group \textit{P2$_{1}$/n} from \textit{Pbnm} \cite{MTAnderson1993}. In present case it is not easy to distinguish between \textit{P2$_{1}$/n} and \textit{Pbnm} symmetries by PXRD because: (i) Mean scattering factor for X-ray from Ni and Mn are nearly equal, (ii) Very small difference in $ \beta $ (the angle between a and c crystallographic axes) for \textit{P2$_{1}$/n} and \textit{Pbnm}. Because of the aforementioned reasons, the superstructure peak (011) at $ 2\theta \sim 9.2^{0} $ in PXRD having intensity 0.6$\%$ of most intense peak at $ 2\theta \sim 15.3^{0} $, is masked in the background contribution. On the other hand, neutron diffraction is able to probe the level of B-site cation ordering in the lattice structure owing to the distinct scattering lengths for Ni and Mn species. Room temperature NPD patterns for SNMO samples illustrated in Figs. \ref{xrdnpdstr}(c, d), show existence of (011) superstructure reflection at $ 2\theta \sim 6.36^{0} $. This confirms dominating cation ordering present in SNMO samples and hence justifies the choice of \textit{P2$_{1}$/n} space group to refine all the PXRD and NPD patterns. Noticeably, the intensity of nuclear superstructure peak (011) with respect to corresponding most intense peak at $ 2\theta \sim 12.36^{0} $ in S$\_$O is larger in comparison to S$\_$D (Figs. \ref{xrdnpdstr}(c, d)), which indicates qualitatively better cation ordering in S$\_$O than S$\_$D. The refined values of lattice parameters, site occupancy factors, atomic coordinates, isotropic thermal coefficients and the reliability indicators are listed in Table \ref{pxrdnpd300krefinetable}. Obtained statistical factors suggest a reasonable match between experimentally observed and simulated pattern. Good agreement between the refined parameters from PXRD and NPD scans indicates the reliability of both diffraction measurements. NPD refinement of Ni and Mn occupancies at \textit{2c} and \textit{2d} sites, reveals presence of $ \sim $3.8(6)$\%$ and 11.8(8)$\%$ ASD in S$\_$O and S$\_$D samples, respectively. As the vacancy defects may influence the electronic and magnetic properties \cite{SMajumder2019}, it is also important to check the possible presence of cation or anion vacancy defects in SNMO system. The scattered X-ray intensity in PXRD measurements depend on electron density of the scattering atom, hence the contributions from heavy Z elements dominate over low Z elements and any attempt to determine atomic concentration of low Z element such as oxygen could be erroneous. In this context, NPD is particularly advantageous as the neutron scattering cross section for oxygen atom is large and significantly different from other constituting elements of SNMO lattice. Therefore, we have utilized Rietveld analysis of NPD profiles measured at T=300 K to estimate oxygen concentration in studied SNMO samples. Obtained occupancy values as presented in Table \ref{pxrdnpd300krefinetable}, suggest that oxygen stoichiometry is maintained in SNMO systems. Empirically calculated values of Goldschmidt tolerance factor \cite{VMGoldschmidt1926} are $ t \approx $ 0.929 for Ni$^{2+} $, Mn$ ^{4+} $ valence states; 0.932 for Ni$ ^{3+} $ low spin (LS), Mn$ ^{3+} $ high spin (HS) states; and 0.923 for Ni$ ^{3+} $ HS, Mn$ ^{3+} $ HS states. The $ t < 1 $ situation leads to tilting of (Ni/Mn)O$ _{6} $ octahedra which causes distortion in the system. 

\begin{figure*}[t]
\centering
\includegraphics[angle=0,width=0.8\textwidth]{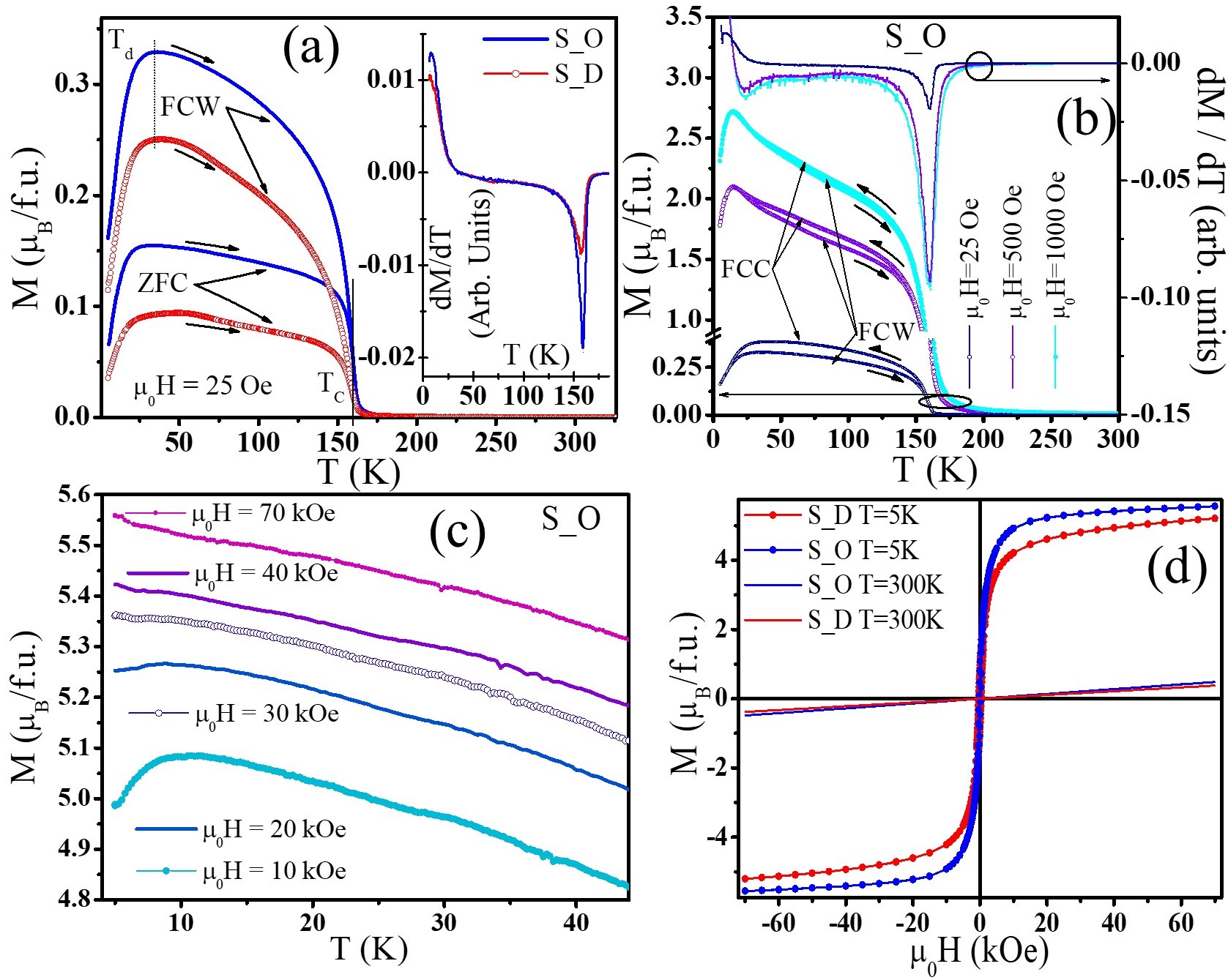}
\caption{(a): Magnetization as a function of temperature M(T) measured under $ \mu_{0} $H=25 Oe for SNMO ordered S$\_$O and disordered S$\_$D samples. Inset show temperature derivative of field cooled warming cycles of M(T). Effect of different measuring magnetic field values on, (b left panel): thermal hysteresis in cooling-warming magnetization cycles, (b right panel): PM to FM transition temperature T$ _{C} $ and (c): low temperature downturn transition T$ _{d} $, examined for S$\_$O sample. (d): Isothermal magnetic hysteresis loops for SNMO samples at T=5 K and 300 K.}\label{mvsth}
\end{figure*}

Absorption spectra across Mn and Ni \textit{K}-edges measured for SNMO samples along with standard references for Mn$ ^{4+} $ (MnO$ _{2} $), Mn$ ^{3+} $ (Mn$ _{2} $O$ _{3} $), Ni$ ^{2+} $ (NiO) and Ni$ ^{3+} $(Ni$ _{2} $O$ _{3} $) are depicted in Figs. \ref{xanes}(a, b). SNMO XANES spectra for both Mn and Ni \textit{K}-edge show characteristic white line just near the absorption edge along with pre-edge features and post-edge oscillatory behavior. XANES spectra at transition metal (TM) \textit{K}-edge is related to photo electron excitation from atomic 1\textit{s} to empty \textit{p}-bands. The pre-edge features in XANES spectra may arise because of 1\textit{s}-3\textit{d} quadrupole or / and weakly allowed dipole transitions. TMO$ _{6} $ octahedral distortions give rise mixed \textit{p-d} states which cause such weakly allowed \textit{s}-\textit{d} transition \cite{FBridges2001}. The multiple scattering contributions from different coordination shells result in post-edge oscillations. It is known that as the valency of TM ion increases, the band edge which is defined as the first inflection point in absorption spectra, shifts towards higher energy values \cite{AHdeVries2003}. Comparing observed \textit{K}-edge XANES spectra (Insets of Figs. \ref{xanes}(a, b)), it can be inferred that: (i) both of the SNMO samples have almost same absorption edge energy for Mn and Ni species, (ii) Mn and Ni ions present in SNMO samples have valency in between 3+, 4+ and 2+, 3+ respectively. Observed mixed valence nature of both TM species indicates charge disproportion between B-site cations through $ Ni^{2+}+Mn^{4+} \longrightarrow Ni^{3+}+Mn^{3+} $, which is also reported in many other double perovskite systems \cite{NSRogadol2005, RIDass2003}. With varying ASD we have not observed apparent changes in spectral characters in XANES, suggesting charge states of constituting elements remain unbiased to ASD densities. 

\begin{figure*}[t]
\centering
\includegraphics[angle=0,width=0.80\textwidth]{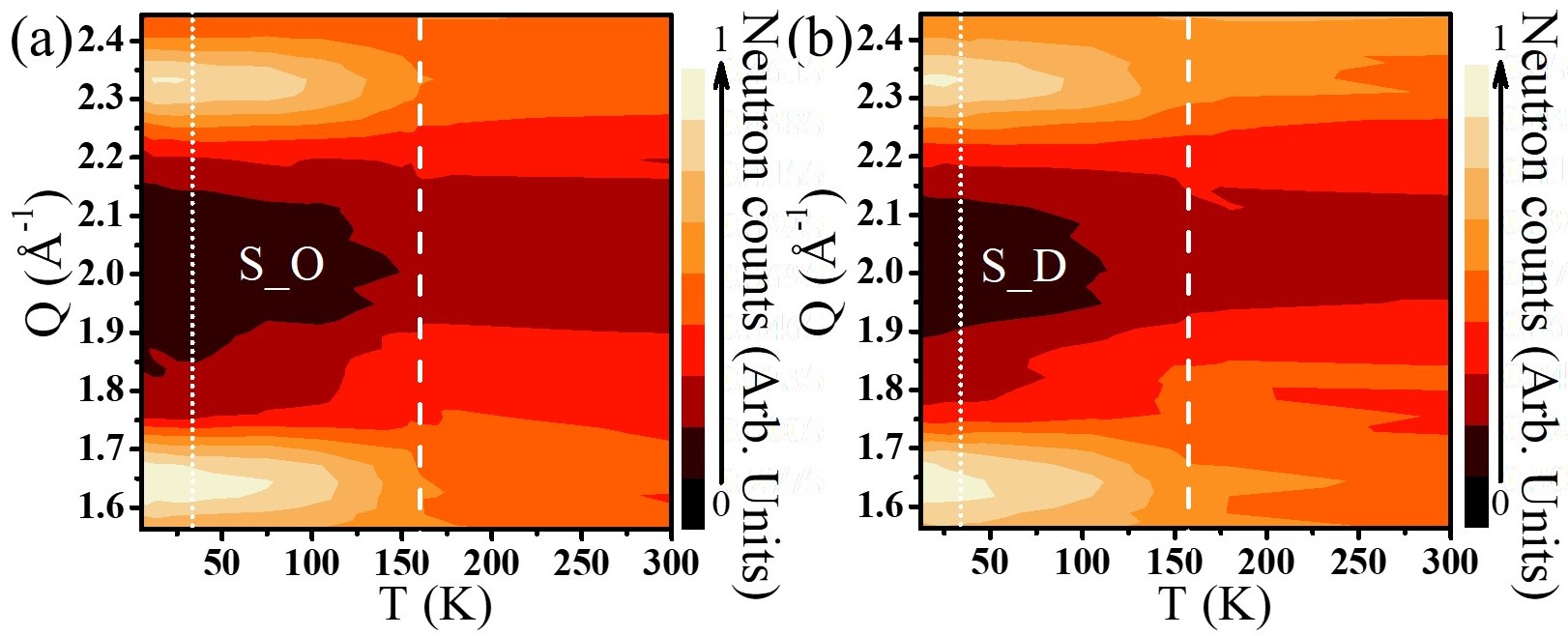}
\caption{Contour map of temperature dependent measured neutron intensity in low Q region for SNMO (a): ordered S$ \_ $O and (b): disordered S$ \_ $D samples. The vertical dashed and dotted lines highlight T$ _{C} $ and T$ _{d} $ respectively, derived from magnetometric measurements.}\label{qcontour}
\end{figure*}
	
The dc magnetometric response acquired for SNMO samples are shown in Fig. \ref{mvsth}. The temperature dependency of magnetic moment M(T) measured in typical zero field cooled (ZFC) warming and field cooled warming (FCW) protocols under 25 Oe of applied magnetic field reveal two distinct magnetic transitions, (i) at T=T$ _{C} $, defined as the onset temperature point of M(T) and (ii) at T=T$ _{d} $, defined as the downturn in M(T). Transition temperature T$ _{C} $ is determined from first temperature derivative of FCW magnetization dM/dT (Inset (i) of Fig. \ref{mvsth}(a)) and found to be T$ _{C} $=159.6 K for S$\_$O and T$ _{C} $=156.6 K for S$\_$D. T$ _{d} $ is estimated from the temperature where M(T) attains the maximum value and approximated to be T$ _{d} \sim $34.1 K for both S$\_$O, S$\_$D samples. Overall magnetic moment decreases with increasing ASD concentrations. In dM/dT curve, a broad inverted cusp like behavior at around T$ \sim $67 K is observed which is more prominent in S$\_$D sample than S$\_$O, as illustrated in Fig. S1 in Supplementary Material (SM) \cite{Supplementalmaterial}. This feature can be attributed to Mn-O-Mn short scale interaction due to anti-site disordered bonds \cite{TKimura2003}. M(T) recorded under different measuring magnetic field values as shown in Fig. \ref{mvsth}(b) for S$\_$O sample, indicate robustness of transition temperature T$ _{C} $. Below T$ _{C} $ a clear thermal hysteresis in between field cooled cooling (FCC) and warming magnetization paths having broad temperature width of $ \bigtriangleup T \sim$140 K in presence of $ \mu_{0} $H=25 Oe is observed for S$\_$O sample as displayed in Fig. \ref{mvsth}(b). With increasing applied field strength ($ \mu_{0} H \gtrsim $1 kOe) the thermal hysteresis vanishes. Similar behavior is also confirmed for S$\_$D sample (data not shown here). The downturn transition shifts towards low temperature side with increasing measuring field values (at $ \mu_{0} $H $ \sim $ 30 kOe) and with a further increase of magnetic field ($ \mu_{0} $H $ > $ 30 kOe) the downturn behavior seems to transform into an upturn as shown in Fig. \ref{mvsth}(c), examined for S$ \_ $O sample. Qualitatively similar trend is observed for S$ \_ $D case (data not shown here) also. The isothermal magnetization M(H) measured at T=5 K presented in Fig. \ref{mvsth}(d), reveals decrease in saturation moment value with increasing ASD in SNMO system. Experimentally estimated moment $ \mu_{expt} $ values at $ \mu_{0}H $=70 kOe, T=5 K are found to be 5.56 $ \mu_{B} $ in S$ \_ $O and 5.21 $ \mu_{B} $ in S$ \_ $D. M(H) at T=300 K show typical paramagnetic (PM) behavior. 

\begin{table*}[]
\centering
\resizebox{\textwidth}{!}{
	\begin{tabular}{ccccccccccccc} 
		\hline
  IRs   & \multicolumn{4}{c}{IR matrices in symbolic form} & \multicolumn{8}{c}{Basis vectors} \\ 
	    & \multicolumn{4}{c}{Symmetry operators} & \multicolumn{8}{c}{Atomic sites} \\
        & 1 & 2 (0,1/2,0) 1/4,y,1/4 & -1 0,0,0 & n (1/2,0,1/2) x,1/4,z & Sm1 & Sm2 & Sm3 & Sm4 & Ni1 & Ni2 & Mn1 & Mn2 \\ 
        \hline\hline
 IR (1) & 1 & 1 & 1 & 1 & (u,v,w) & (-u,v,-w) & (u,v,w) & (-u,v,-w) & (r,s,t) & (-r,s,-t) & (o,p,q) & (-o,p,-q) \\
 IR (2) & 1 & 1 & -1 & -1 & (u,v,w) & (-u,v,-w) & (-u,-v,-w) & (u,-v,w) &   &   &   &   \\ 
 IR (3) & 1 & -1 & 1 & -1 & (u,v,w) & (u,-v,w) & (u,v,w) & (u,-v,w) & (r,s,t) & (r,-s,t) & (o,p,q) & (o,-p,q) \\
 IR (4) & 1 & -1 & -1 & 1 & (u,v,w) & (u,-v,w) & (-u,-v,-w) & (-u,v,-w) &   &   &   &   \\ 
		\hline\hline
	\end{tabular}
}
	\caption{SNMO magnetic structure character table generated for \textit{P2$_{1}$/n} space group with \textbf{k}=(0, 0, 0) propagation vector.}
	\label{tableIR}	
	\end{table*} 

\begin{figure*}[t]
\centering
\includegraphics[angle=0,width=0.85\textwidth]{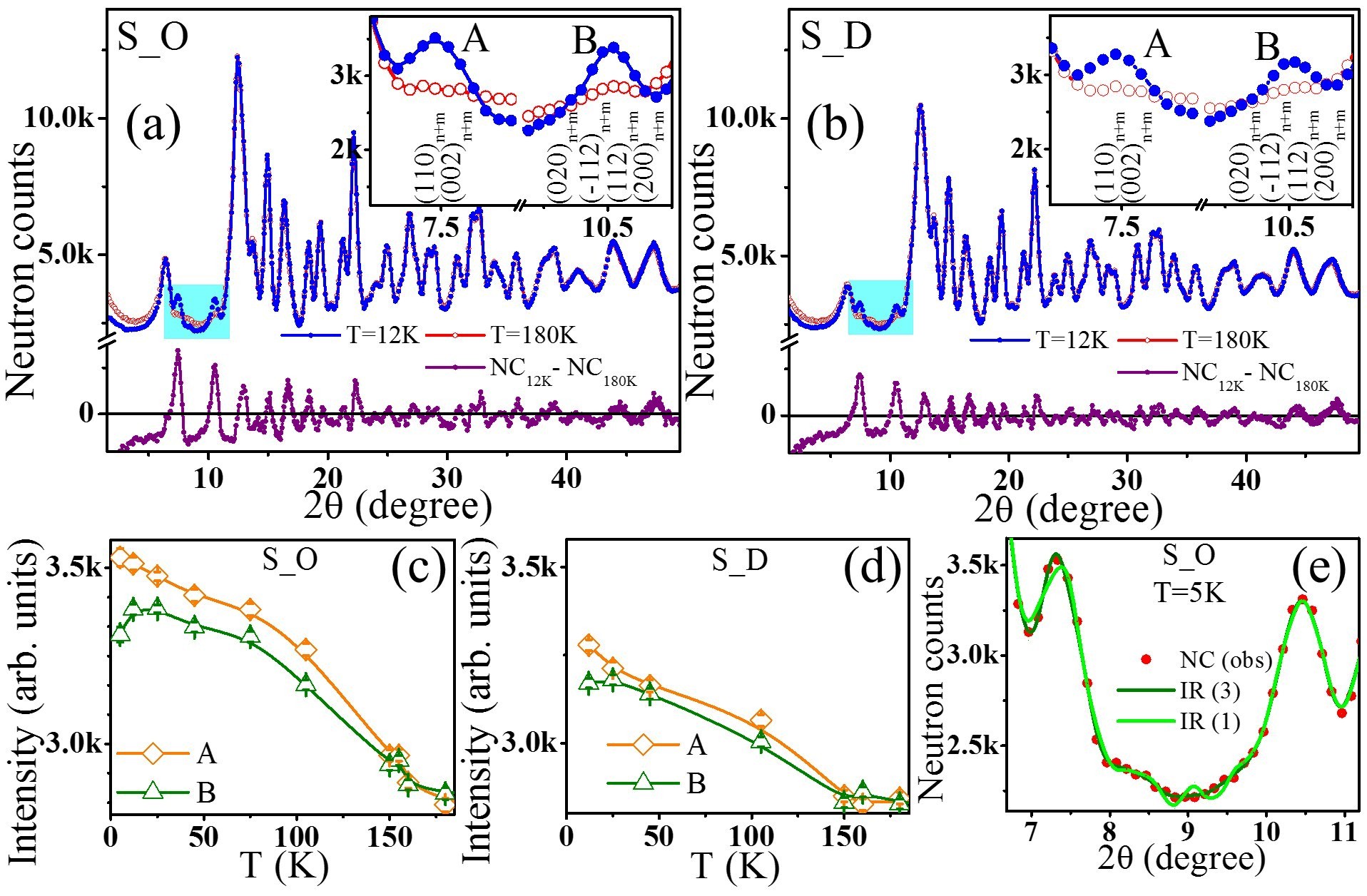}
\caption{(a, b): NPD patterns collected at T=12 K, 180 K along with corresponding temperature subtracted difference spectra for SNMO ordered S$ \_ $O and disordered S$ \_ $D samples. Insets of (a, b), show enlarged view of low angle peaks `A': (110)$_{n+m}$+(002)$_{n+m}$ and `B': (020)$_{n+m}$+(112)$_{n+m}$+(200)$_{n+m}$, where n, m corresponds to nuclear and magnetic contributions respectively. Temperature dependency of NPD intensity at features `A' and `B' 2$ \theta $ positions for (c): S$ \_ $O and (d): S$ \_ $D samples. Open symbols are for observed data and solid lines are for guide to the eyes. (e): Best fitted curves along with experimentally observed NPD diffractograms at T=5 K across `A' and `B' reflections modeled with different IRs for S$ \_ $O sample.}\label{npdtdiff}
\end{figure*}

Figures \ref{qcontour}(a, b) depict the contour maps of temperature  dependent neutron scattering intensity in low Q regime measured for SNMO samples. Transition temperatures obtained from aforementioned bulk magnetometric results are marked by vertical dashed (T$ _{C} $) and dotted (T$ _{d} $) lines. With change in temperature, distinguishable variations in measured neutron counts are observed at Q $ \sim $1.6 and 2.3 $ \AA^{-1} $ in the vicinity of T$ _{C} $ and T$ _{d} $ transitions. In order to have better visualization, measured temperature dependent NPD profiles are vertically stacked as shown by Figs. S2(a, b) in SM \cite{Supplementalmaterial}. The comparison of NPD patterns recorded at T=12 K (T$ < $T$ _{C} $) and T=180 K (T$ > $T$ _{C} $) for SNMO samples are highlighted in Figs. \ref{npdtdiff}(a, b). Point by point difference spectra between these two temperatures are evaluated to eliminate the scattering contributions from nuclear structure and background. Temperature dependent PXRD analysis (not shown here) confirms the absence of symmetry related structural phase transition within the investigated temperature region (5 K$ \leqslant $T$ \leqslant $300 K). Therefore, the difference patterns present, magnetic scattering along with temperature driven effect (Debye) on nuclear reflections. The negative values in difference spectra arise from dominating diffuse scattering contributions at T=180 K superimposed underneath the Bragg's reflections. For both SNMO samples, no new magnetic peak emerges. Magnetic reflections are contributed by means of intensity enhancement at same nuclear Bragg's peak positions. Hence, in SNMO the magnetic and nuclear unit cells coincide and \textbf{k}=(0, 0, 0) is chosen as propagation vector for magnetic structure. Propagation vector \textbf{k}=(0, 0, 0) is also used to define magnetic unit cell in other members of ANMO family \cite{SanchezBenitez2011, MRetuerto2015, HNhalil2015}. The appearance of magnetic peaks at same nuclear Bragg's positions indicates FM nature of SNMO samples. At lower $ 2\theta $ values of NPD profile distinguishable changes are prominently observed in peak intensity against temperature variation. Such abrupt change in peak intensity with temperature suggests that these low angle reflections have dominating magnetic characters over nuclear contribution. The peak at $ 2\theta \sim$ 7.44 $ \AA $ marked as `A' is indexed by closely spaced Bragg's reflections $ (110)_{n+m} $ and $ (002)_{n+m} $, where n and m corresponds to nuclear and magnetic contributions, respectively. The reflection at $ 2\theta \sim$ 10.57 $ \AA $ marked as `B' is indexed by closely spaced $ (020)_{n+m} $, $ (112)_{n+m} $ and $ (200)_{n+m} $ Bragg's peaks. The thermal evolution of features A and B as shown in Figs. \ref{npdtdiff}(c, d) behave like magnetic order parameter. On lowering temperature both A and B features reveal onset in intensity at around T=T$ _{C} $. On further decrease of temperature at T$ \sim $T$_{d}$ intensity of feature B encounters a downturn whereas intensity of A increases. In a previous study \cite{G R Haripriya2011}, cation disordered (\textit{Pbnm} symmetry) Ho$ _{2} $FeCoO$ _{6} $ double perovskite is found to show significant intensity redistribution in magnetic Bragg's lines at low temperature, which is identified as spin reorientation transition leading to transformation from mixed magnetic structure to single magnetic structure. However, this is not the case for SNMO system, as here single magnetic structure is evidenced for all temperature values below T=T$ _{C} $ (discussed later). Therefore, the observed magnetic intensity redistribution indicates temperature driven changes in magnetic structure of SNMO samples. 

\textit{P2$_{1}$/n} space group symmetry allowed possible magnetic structures for \textbf{k}=(0, 0, 0) propagation vector are computed by the help of representation theory analysis developed by Bertaut \cite{EBertaut1962, EBertaut1968}. Obtained IRs and associated basis vectors for the Sm, Ni and Mn sublattices are listed in Table \ref{tableIR}. Different trial magnetic representations are tested in refinement model and corresponding quality of fitting are checked. As presented in Fig. \ref{npdtdiff}(e), the best fit to experimental pattern is obtained for IR (3). 

\begin{figure*}[t]
\centering
\includegraphics[angle=0,width=0.7\textwidth]{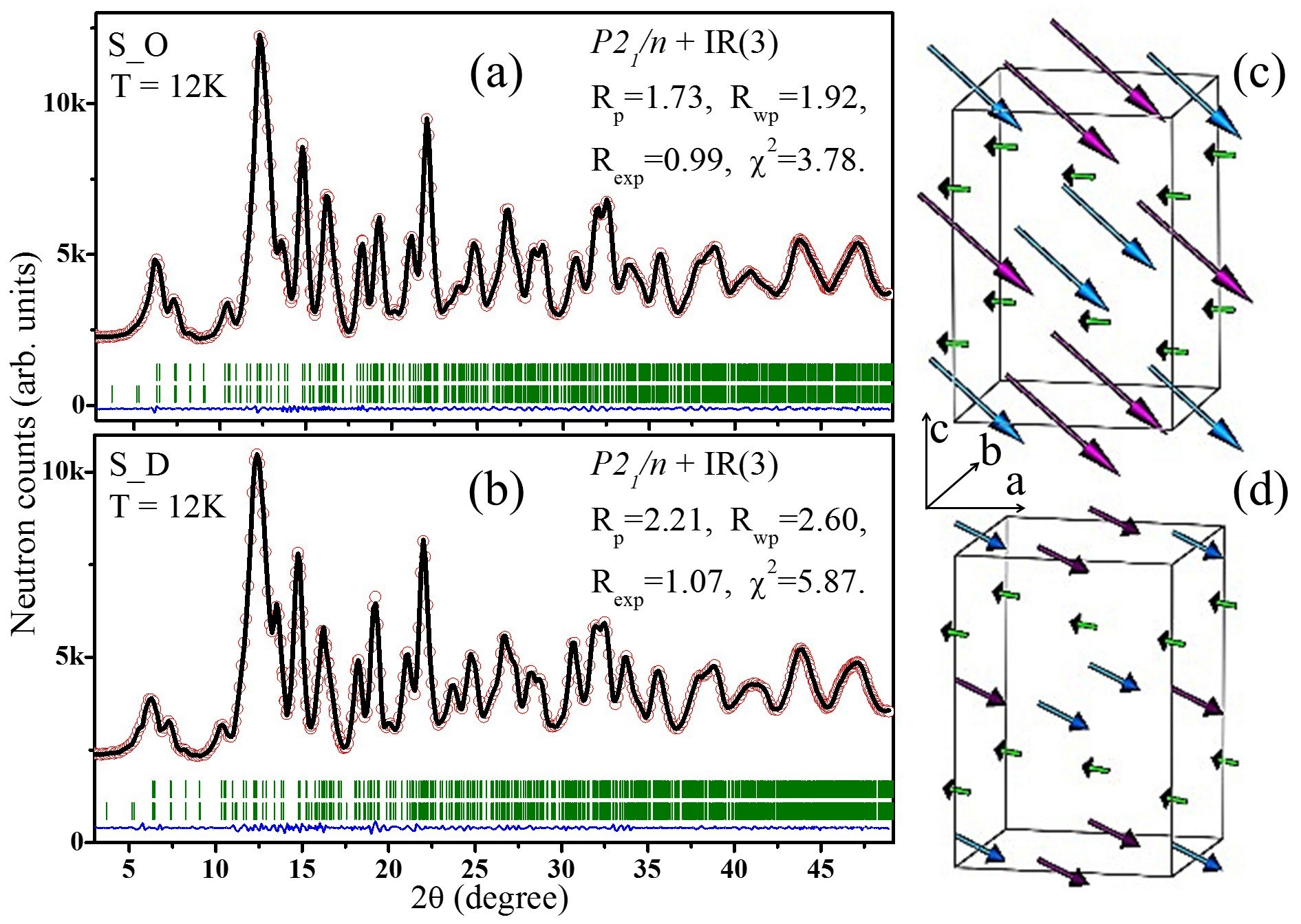}
\caption{(a, b): Rietveld analysis of NPD $ 2\theta $ scans recorded at T=12 K showing observed (red open circles), calculated (black solid line) and difference (blue solid line) pattern along with Bragg's positions (green vertical bars) for SNMO ordered S$ \_ $O and disordered S$ \_ $D samples. The Rietveld generated patterns are simulated with monoclinic \textit{P2$_{1}$/n} (SG 14) nuclear structure and IR (3) magnetic structure with \textbf{k}=(0, 0, 0) propagation vector. The goodness of fitting indicators, for S$ \_ $O data: R$_{p}$=1.73, R$_{wp}$=1.92, R$_{exp}$=0.99, $\chi^2$=3.78 and for S$ \_ $D data: R$_{p}$=2.21, R$_{wp}$=2.60, R$_{exp}$=1.07, $\chi^2$=5.87. Refined magnetic structures at T=12 K for (c): S$ \_ $O and (d): S$ \_ $D samples.}\label{magstr12k}
\end{figure*}

\begin{figure*}[t]
\centering
\includegraphics[angle=0,width=0.98\textwidth]{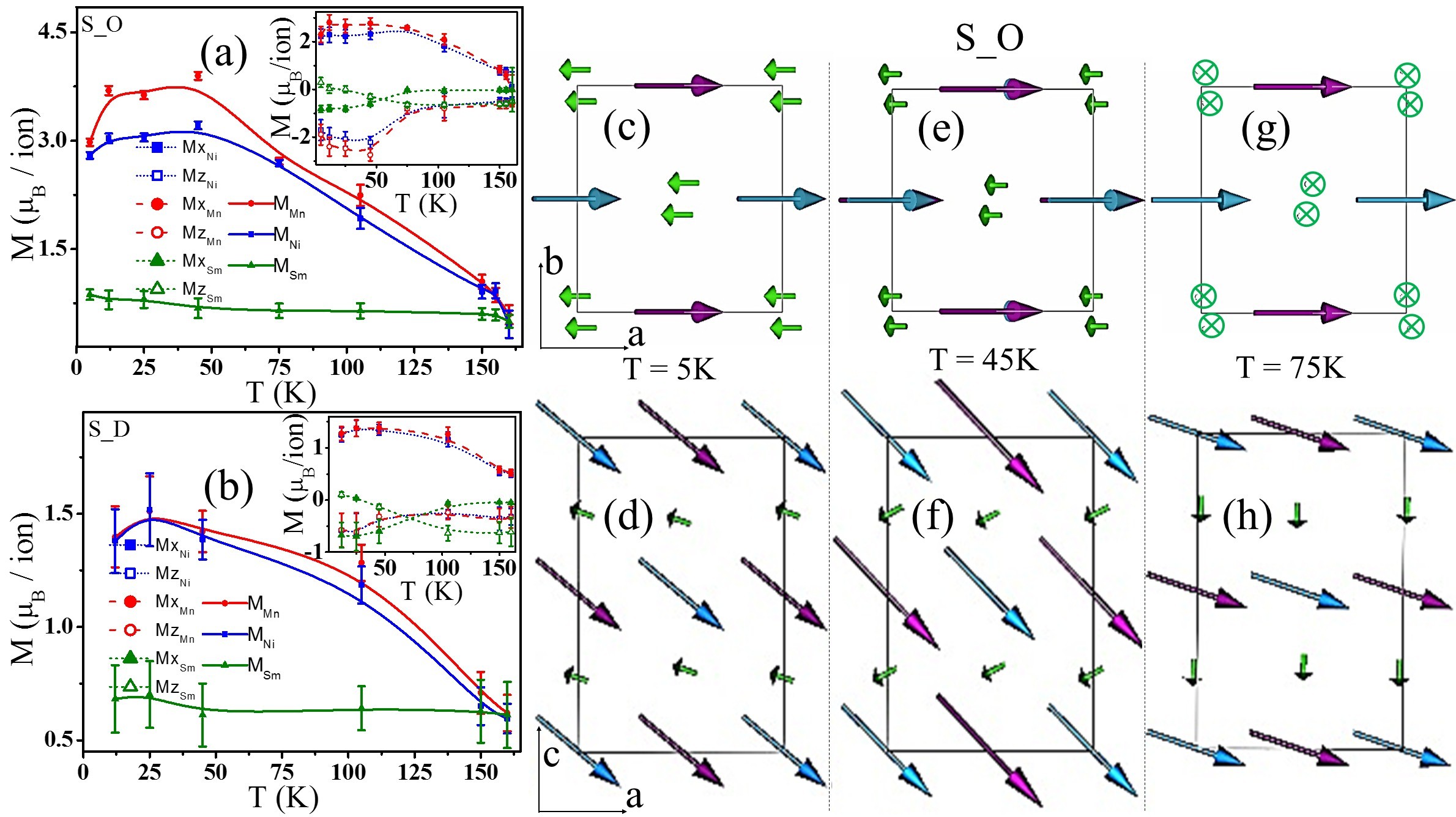}
\caption{Temperature dependency of element specific sublattice moments for SNMO (a): ordered S$ \_ $O and (b): disordered S$ \_ $D samples. Insets of (a, b): Variation of moment components along x (a-axis) and z (c-axis) directions. Geometric symbols are for values obtained from refinements and solid lines are for guide to the eyes. Spin arrangement for S$ \_ $O sample in (c, e, g): a-b plane and (d, f, h): a-c plane, obtained by refining NPD data recorded at T=5 K, 45 K and 75 K, respectively. The Rietveld fitting of NPD $ 2\theta $ scans recorded at T=5 K, 45 K and 75 K are presented in Fig. S3 in SM \cite{Supplementalmaterial}. Goodness of fitting indicators, for T=5 K data are R$_{p}$=2.03, R$_{wp}$=2.23, R$_{exp}$=1.45, $\chi^2$=2.37; for T=45 K data: R$_{p}$=1.59, R$_{wp}$=1.77, R$_{exp}$=0.92, $\chi^2$=3.73 and for T=75 K data are R$_{p}$=1.72, R$_{wp}$=1.94, R$_{exp}$=0.98, $\chi^2$=3.91.}\label{magstrvst}
\end{figure*}

Rietveld refinements of NPD data at T=12 K, presented in Figs. \ref{magstr12k}(a, b), show good agreement between experimental and calculated patterns. Obtained spin arrangements at T=12 K, as illustrated in Figs. \ref{magstr12k}(c, d), reveal $ F_{x}F_{z} $ magnetic structure of Ni-Mn network. Moment vectors are found to lie in a-c plane. Nonzero moment component along b-axis does not improve the fitting quality and even it diverges. Therefore, magnetic structure analysis unambiguously confirms that below T=T$ _{C} $ SNMO undergoes collinear FM ordering of Ni, Mn magnetic sublattices. Thermal variation of element specific magnetic moments from Sm, Ni and Mn sublattices are presented in Figs. \ref{magstrvst}(a, b). The temperature evolution of magnetic moment components along a (x) and c (z) directions obtained from Rietveld refinement are shown in Insets of Figs. \ref{magstrvst}(a, b). With decreasing temperature, Ni and Mn spins develop long range ordering in FM configuration and corresponding sublattice magnetic moment values start increasing below T=T$ _{C} $. We have not observed any signature of Sm sublattice long range ordering within 5 K$ \leq $T$ \leq $300 K. However, PM Sm moments are polarized due to the internal exchange field from Ni-Mn ordered sublattice. Upon further cooling across T=T$ _{d} $ polarized Sm moments align opposite to Ni-Mn moments, as illustrated in Figs. \ref{magstrvst}(c-h). With increasing ASD concentration the magnetic moment values are found to decrease as depicted in Figs. \ref{magstrvst}(a, b). This trend is in agreement with magnetometric observation.

\begin{figure*}[t]
\centering
\includegraphics[angle=0,width=0.8\textwidth]{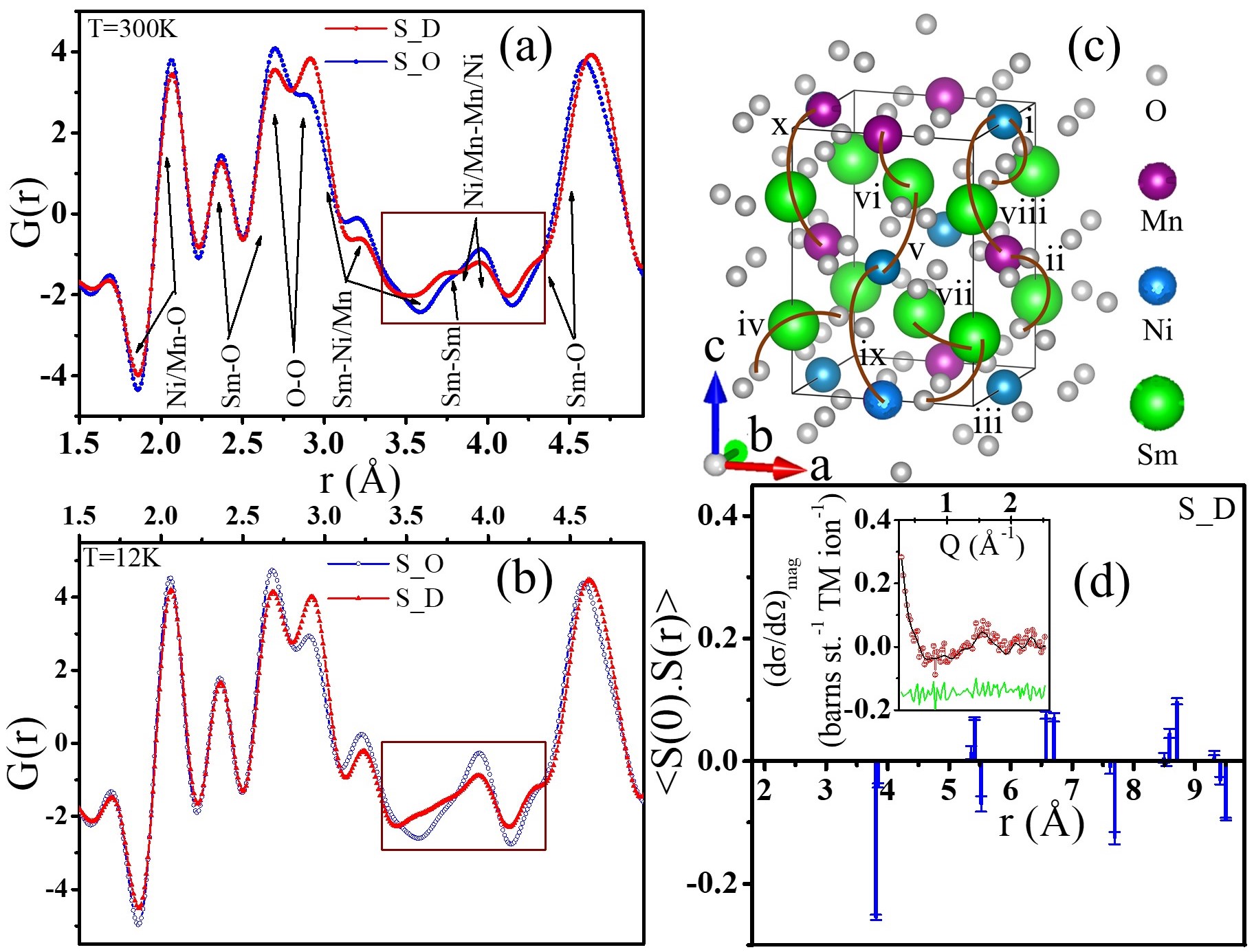}
\caption{Real space PDF histograms at (a): T=300 K and (b): T=12 K for SNMO ordered S$ \_ $O and disordered S$ \_ $D samples. Feature for Ni/Mn-Mn/Ni atomic pairs are highlighted by box. (c): Refined nuclear structure for SNMO system, demonstrating different inter-atomic separation distances in $ \AA $; (i) Ni-O1: 1.819(4), Ni-O2: 1.960(3), Ni-O3: 1.968(4), (ii) Mn-O1: 2.132(4), Mn-O2: 1.938(3), Mn-O3: 1.995(4), (iii) Sm-O1 short: 2.393(3), long: 4.221(6), 4.493(5), 4.504(6), Sm-O2 short: 2.559(5), long: 4.415(5), 4.437(6), 4.578(5), Sm-O3 short: 2.576(6), long: 4.328(6), 4.486(5), 4.539(5), (iv) O1-O2: 2.638(5), O1-O3: 2.924(5), (v) Sm-Ni: 3.106(4), 3.257(5), 3.345(5), 3.613(4), (vi) Sm-Mn: 3.105(4), 3.257(5), 3.348(5), 3.612(4), (vii) Sm-Sm: 3.771(7), (viii-x) Ni/Mn-Mn/Ni: 3.8093(4), 3.8470(3). Here, number in the parentheses represent the error bars on the last digit. Ni-Mn pair belongs to B-site cation ordered structures whereas Ni-Ni and Mn-Mn pairs are associated with cation disordered structures. (d): Radial spin-spin correlation distribution obtained from fitting diffuse magnetic scattering data (Inset) at T=180 K for S$ \_ $D sample. In Inset red open circles are for observed, black solid line is for calculated and green solid line is for difference pattern.}\label{npdfspincorrel}
\end{figure*}

Sine Fourier transformation of corrected neutron total scattering intensity S(Q) is calculated as \cite{TEgami1991, MPasciak2019}, 
\begin{equation}\label{EqGr}\nonumber
G(r) = K (2/\pi) \int_{Q_{min}}^{Q_{max}} [S(Q)-1]Q{\rm sin}(Qr)dQ
\end{equation} 
where $ r $ is atomic pair separation distance, $ K $ is absolute scale normalization factor and S(Q) has been normalized to 1 at high Q after subtraction of incoherent scattering from the diffraction pattern. Figs. \ref{npdfspincorrel}(a, b) present a comparison of SNMO samples having different ASDs, in low $ r $ region of calculated real space radial distribution histograms at T=300 K and 12 K, respectively. PDF is an ideal probe for local structural modifications and cationic arrangement because it considers both Bragg's and diffuse scattering on equal footage. In nuclear PDF profile, the peak positions contain information about all possible atom-atom coordination distances. Ideally, each atom pair should contribute to well define delta function. Whereas in real system, distribution of inter-atomic separations due to disorder causes peak broadening. The peak heights depend on scattering factors of atomic pairs. The integrated intensity of a given peak corresponds to coordination numbers of atom pair \cite{TEgami2012, DHou2018}. At T=300 K, as SNMO is PM (Fig. \ref{mvsth}(d)), the dominating contribution in observed PDF comes from nuclear (Bragg's + diffuse) scattering. Different regions of these PDF are identified from refined nuclear structure displayed in Fig. \ref{npdfspincorrel}(c). Noticeable differences are observed in PDF histograms of SNMO samples with different ASD densities across some specific regions, where B-site (Ni/Mn) atoms are involved. Particularly, the feature around $ r \sim $3.9 $\AA$ which corresponds to Ni/Mn-Mn/Ni atomic pair contribution, becomes wider in S$ \_ $D than S$ \_ $O, as highlighted in Figs. \ref{npdfspincorrel}(a, b). More peak broadening again confirms the presence of more ASD in S$ \_ $D as compared to S$ \_ $O. With decreasing temperature PDF features become more prominent and narrower, as shown in Figs. S4(a, b) in SM \cite{Supplementalmaterial}. This is because, in low temperature PDF, (i) pair correlations are strong as magnetic (Bragg's + diffuse) scattering also has significant contribution along with nuclear scattering and (ii) with reducing temperature the amplitude of thermal vibrations in atom pairs is reduced. Comparison of PDF at T=12 K as presented in Fig. \ref{npdfspincorrel}(b), shows much more broadening than T=300 K case for Ni/Mn-Mn/Ni pair correlation features with increasing ASD. This again indicates weaker magnetic long range (Bragg's) and stronger diffuse (nuclear+magnetic) scattering due to more ASD present in S$ \_ $D sample.

We have already observed the signature of diffuse magnetic scattering contributions superimposed underneath the Bragg's reflections at T=180 K (Figs. \ref{npdtdiff}(a, b)). In order to extract diffuse magnetic scattering contribution, NPD pattern collected at T=300 K is subtracted from T=180 K pattern as presented in Inset of Fig. \ref{npdfspincorrel}(d). This temperature subtraction eliminates nuclear Bragg's scattering contribution from T=180 K data. Cation ordered SNMO lattice can be considered as alternate stacking arrangement of SmNiO$ _{3} $ and SmMnO$ _{3} $ single perovskites. Whereas, in cation disordered SNMO lattice, SmNiO$ _{3} $ and SmMnO$ _{3} $ have random distribution. Therefore, in SNMO system, magnetic interactions are possible in between Sm-Sm, Sm-Ni, Sm-Mn, Ni-Ni, Ni-Mn and Mn-Mn spin pairs via intermediate O ions. Due to long separation between cation pairs, superexchange through intermediate O ions dominate over direct exchange interaction between spins. Generally rare earth orders at very low temperature (as here, no signature of Sm ordering within 5 K$ \leqslant $T$ \leqslant $300 K) and has relatively very small moment (as observed from Figs. \ref{magstrvst}(a, b)) values in comparison to TM ions. Hence in diffuse magnetic scattering study, interactions from spin pairs involving Sm ions can be neglected. The long range FM ordering of Ni-Mn sublattices in SNMO are observed below T$_{C} \simeq $160 K. In SmMnO$ _{3} $ \cite{TKimura2003} and SmNiO$ _{3} $ \cite{MLMedarde1997} perovskites, Mn-Mn and Ni-Ni moments are reported to order in AFM configurations, below T$ _{N} \simeq $ 59 K and 225 K, respectively. Therefore, the temperature subtracted scattering intensity at T=180 K, should have dominating contribution from Ni-Ni spin pairs. RMC approach on 10$ \times $10$ \times $10 super-cell filled with randomly oriented classical spin vectors (4000 spins) at B-site sublattices obeying periodic boundary conditions, is employed to analyze diffuse magnetic scattering. Spin rotation at random site is performed following algorithm acceptance criteria until the best possible fitting of observed data is achieved. RMC fit of diffuse magnetic scattering data using Ni magnetic form factor and obtained real space spin-spin correlation distribution at T=180 K are presented in Fig. \ref{npdfspincorrel}(d). Here the radial spin-spin correlation distribution function $ \langle S(0).S(r) \rangle $ represents average scalar product of spin pairs situated at center $ r $=0 and at a distance $ r $. The negative value of spin correlation function at nearest neighbor site ($ r \sim$ 3.8 $ \AA $) unambiguously confirms the presence of short scale Ni-Ni AFM interactions in SNMO. Such short scale coupling originates from ASD Ni-O-Ni bonds in SNMO.

In present scenario, due to the coexisting of ordered, disordered phases and mixed valency of B-site cations a number of magnetic exchange interactions are possible in SNMO system. (i) FM super exchange interaction through Ni$ ^{2+} $(e$ ^{2}_{g} $)-O-Mn$ ^{4+} $(e$ ^{0}_{g} $) in 180$ ^{0} $ geometry from cation ordered configurations \cite{Goodenoughl1995, Kanamoril1959}. (ii) AFM interaction because of virtual charge transfer between Ni$ ^{2+} $(e$ ^{2}_{g} $)-O-Ni$ ^{2+} $(e$ ^{2}_{g} $) and Mn$ ^{4+} $(t$ ^{3}_{2g} $)-O-Mn$ ^{4+} $(t$ ^{3}_{2g} $) ASD pairs. (iii) FM interaction due to vibronic superexchange between Ni$ ^{3+} _{LS} $(e$ ^{1}_{g} $)-O-Mn$ ^{3+} _{HS} $(e$ ^{1}_{g} $) (where LS: low spin state, HS: high spin state) cation ordered pairs. (iv) AFM superexchange interaction between ASD pairs via Ni$ ^{3+} _{LS} $(e$ ^{1}_{g} $)-O-Ni$ ^{3+} _{LS} $(e$ ^{1}_{g} $) and Mn$ ^{3+} _{HS} $(e$ ^{1}_{g} $)-O-Mn$ ^{3+} _{HS} $(e$ ^{1}_{g} $). (v) Exchange interaction between all combinations of Ni and Mn pair, when Ni$^{3+}$ stabilizes in HS state \cite{FFan2016}. (vi) AFM superexchange interaction in 180$ ^{0} $ linkage between Ni$ ^{2+} $(e$ ^{2}_{g} $)-O-Ni$ ^{3+} $(e$ ^{1}_{g} $). (vii) FM double exchange interaction between Mn$ ^{3+} $(e$ ^{1}_{g} $)-O-Mn$ ^{4+} $(e$ ^{0}_{g} $). In addition, these aforementioned magnetic exchange interactions can be affected by local octahedral distortion, since both Ni$ ^{3+} $ and Mn$ ^{3+} $ have orbital degeneracy and are Jahn-Teller active. As observed in present case, SNMO samples have significant amount of cation ordered phase (cation ordering in S$ \_ $O is $ \sim $96.1$ \% $ and in S$ \_ $D is $ \sim $86.5$ \% $) and hence magnetic interactions (i) and (iii) have dominating contribution. However, the Ni$ ^{3+} _{LS} $-O-Mn$ ^{3+} _{HS} $ vibronic superexchange interaction (iii) is less stable in comparison with static superexchange interaction (i) through Ni$ ^{2+} $-O-Mn$ ^{4+} $. With increasing ASD densities, contributions from (ii) and (iv) interactions enhance. The phase boundary effects through interactions (vi) and (vii) are relatively feeble in comparison with contributions from ordered or disordered regions. As a consequence of these coexisting exchange pathways, the magnetic behavior of SNMO comprises of competing FM-AFM interactions, which are respectively originated from the B-site ordered and disordered structures. Observed transition at T=T$ _{C} $ in SNMO system is assigned to long range FM ordering from Ni-O-Mn B-site ordered chains. The theoretical effective magnetic moments for different spin configurations defined as, $ \mu_{theo} = \sqrt{ 2\times\mu_{Sm}^{2}+\mu_{Ni}^{2}+\mu_{Mn}^{2} } $, where $ \mu_{Sm} = g \sqrt {J(J+1)}  $ with $g \simeq $ 0.2857 and $ \mu_{{Ni/Mn}} = g \sqrt {S(S+1)}$ with $g \simeq $ 2.00, yield values:
\begin{center}
4.94 $ \mu_{B} $ for \\ \begin{normalsize}
[Sm$^{3+}$ (J=5/2), Ni$ ^{2+}$ (S=1), Mn$ ^{4+}$ (S=3/2)],
\end{normalsize} \\
6.36 $ \mu_{B} $ for \\ \begin{normalsize}
[Sm$^{3+}$ (J=5/2), Ni$ ^{3+}_{HS}$ (S=3/2), Mn$ ^{3+}_{HS}$ (S=2)],
\end{normalsize} \\ 
and 5.33 $ \mu_{B} $ for \\ \begin{normalsize}
[Sm$^{3+}$ (J=5/2), Ni$ ^{3+}_{LS}$ (S=1/2), Mn$ ^{3+}_{HS}$ (S=2)].
\end{normalsize}
\end{center} 
Therefore, observed $ \mu_{expt} $ values suggest mixed valence spin interaction of B-site cations. XANES measurements for SNMO samples suggest that mixed valence nature of both Ni and Mn ions remain unbiased to ASD concentration. Therefore, the smaller $ \mu_{expt} $ value of S$ \_ $D than S$ \_ $O is solely due to different level of cation ordering. ASD mediated short scale AFM couplings through Ni-O-Ni or Mn-O-Mn bond pairs weaken the predominant FM interactions in ordered host matrix. Consequently, increase of ASD concentration results in reduction of net magnetic moment and saturation magnetization. This decrease of moment values with increasing ASD concentration is confirmed from magnetometric measurements as well as NPD studies. With increasing ASD density short range correlations
increase in the system and this leads more prominent inverted cusp like broad transition in S$ \_ $D sample as compared to S$ \_ $O case. The thermo-magnetic irreversibility in FCC and FCW magnetization cycles in any system can be due to magnetic frustration \cite{RPradheesh2012} or first order phase transition \cite{MTripathi2017} or spin glass like state or domain wall pinning or arrested dynamic behavior \cite{RPMadhogaria2020}. In present case, observed thermal hysteresis in SNMO samples are possibly originated from frustration due to competing exchange interactions in mixed magnetic phases. At high measuring field values ($ \mu_{0} H \gtrsim $1 kOe) the long range magnetic interaction from ordered phase dominates over the diluted short interaction in disordered phase and consequently the thermo-magnetic irreversibility vanishes. The sharp reduction in magnetization below T=T$ _{d} $ suggests the possibility of anti-parallel alignment of constituent moments in the magnetic lattice. During cooling from high temperature PM state below T$ _{C} $, Ni and Mn ions order in FM configuration and impose an internal magnetic field H$ _{int} $ at Sm sites. Experiencing the internal exchange field effect exerted by long range ordered Ni-Mn sublattice, PM Sm moments are polarized in anti-parallel direction with respect to Ni-Mn network. The opposite alignment of polarized rare-earth moments with respect to TM moments due to the presence of an internal magnetic field has been observed in some perovskite and double perovskite structures \cite{SanchezBenitez2011, MTripathi2019, SPal2019}. With decreasing temperature, there is increase of orientational ordering of polarized Sm PM moments. At T=T$ _{d} $ the competition between oppositely aligned overall magnetic moment of Ni-Mn sublattice and polarized PM moment of Sm site results in downturn behavior in temperature dependent magnetization. When the measuring field balances the internal field (therefore, H$_{int} \simeq$ 30 kOe) there is no longer any decrease in magnetic moment with decreasing temperature. 

\section{CONCLUSION}
In summary, we have addressed the microscopic magnetic configuration of Sm$ _{2} $NiMnO$ _{6} $ double perovskite system. Depending on the calcination route, SNMO samples crystallize with differing cation arrangements but in the same monoclinic \textit{P2$_{1}$/n} structure. We have confirmed that the mixed valence nature of both Ni and Mn species arising due to $ Ni^{2+}+Mn^{4+} \longrightarrow Ni^{3+}+Mn^{3+} $ charge disproportion, is independent with respect to variation of anti-site disorder concentration. High energy ($ \lambda \sim $0.5 $ \AA $) neutrons are utilized to overcome the high neutron absorption of natural Sm ions and to record the thermal variation of diffraction patterns. The degree of anti-site disorders is quantified by refining the intensity corresponding to nuclear Bragg's reflections. Below the magnetic ordering temperature (T$ < $160 K), the moments at Ni and Mn sites follow long range collinear ferromagnetic arrangement with spin components along a and c crystallographic axes, irrespective of different anti-site disorder extents. Thermal evolution of individual ionic magnetic moments indicates the opposite alignment of polarized Sm moments with respect to Ni-Mn sublattice moments below T $ \sim $35 K. Pair distribution function calculations depict the decrement in magnetic Bragg's intensity, whereas the enhancement in diffuse (nuclear+magnetic) scattering with increasing the degree of anti-site disorders. Reverse Monte Carlo analysis of diffuse magnetic scattering profiles reveal the existence of anti-site disorder mediated Ni-Ni short range antiferromagnetic couplings ranging up to first nearest neighbor distances, which sustain even above ferromagnetic ordering temperature. As a consequence of these additional Ni-O-Ni (or Mn-O-Mn) local antiferromagnetic interactions in predominantly ordered Ni-O-Mn ferromagnetic host matrix, the magnetic behavior of SNMO comprises of coexisting magnetic phases in wide temperature regime. The present study will in general, provide a route map to control the characteristic magnetic aspects by proper tuning the anti-stie disorders in all the prototypical double perovskite systems.     
\\
\section*{ACKNOWLEDGMENTS}
Authors gratefully acknowledge Dr. Tapan Chatterji (ILL, France) for fruitful discussions regarding NPD measurements. Thanks to ILL, France; Elettra-Sincrotrone, Italy and Indus Synchrotron RRCAT, India for giving access to experimental facilities. Authors acknowledge the Department of Science and Technology (DST), Government of India and Jawaharlal Nehru Centre for Advanced Scientific Research for providing financial support through DST-Synchrotron-Neutron project (No. JNC/Synchrotron and Neutron/2019-20/In-009) to perform experiments at ILL. Authors are thankful to DST, Indian Institute of Science, Italian Government and Elettra for providing financial support through Indo-Italian Program of Cooperation (No. INT/ITALY/P-22/2016 (SP)) to perform experiments at Elettra-Sincrotrone. Thanks to Dr. Karin Schmalzl (Forschungszentrum Juelich, Germany) for his help in NPD experiments and Dr. Rajamani Raghunathan (UGC-DAE CSR, India) for fruitful discussions. Authors acknowledge Mr. M. N. Singh (RRCAT, India) for his skillful technical support in PXRD measurements.

\bibliography{}

\begin{thebibliography}{999}

\bibitem{ABalandin2011}
A. Balandin, 
Nat. Mater \textbf{10}, 569-581 (2011).

\bibitem{RNoriega2013}
R. Noriega, J. Rivnay, K. Vandewal, F. P. V. Koch, N. Stingelin, P. Smith, M. F. Toney and A. Salleo, 
Nat. Mater \textbf{12}, 1038-1044 (2013).

\bibitem{JBowles2013}
J. Bowles, M. Jackson, T. Berqu\'{o}, P. A. S{\o}lheid and J. S. Gee, 
Nat. Commun. \textbf{4}, 1916 (2013).

\bibitem{MTAnderson1993}
M. T. Anderson, K. B. Greenwood, G. A. Taylor, K. R. Poeppelmeier, 
Prog. in Solid St. Chem, \textbf{22}, 3, 197-233 (1993).

\bibitem{MGHernandez2001}
M. Garc\'{i}a-Hern\'{a}ndez, J. L. Mart\'{i}nez, M. J. Mart\'{i}nez-Lope, M. T. Casais, and J. A. Alonso, 
Phys. Rev. Lett. \textbf{86}, 2443 (2001).

\bibitem{DDSarma2001}
D. D. Sarma, S. Ray, K. Tanaka, M. Kobayashi, A. Fujimori, P. Sanyal, H. R. Krishnamurthy, and C. Dasgupta, 
Phys. Rev. Lett. \textbf{98}, 157205 (2007).

\bibitem{DChoudhury2012}
D. Choudhury, P. Mandal, R. Mathieu, A. Hazarika, S. Rajan, A. Sundaresan, U. V. Waghmare, R. Knut, O. Karis, P. Nordblad, and D. D. Sarma, 
Phys. Rev. Lett \textbf{108}, 127201 (2012).

\bibitem{NSRogadol2005}
N. S. Rogado, J. Li, A. W. Sleight, M. A. Subramanian, 
Adv. Mater., \textbf{17}: 2225-2227 (2005).

\bibitem{YShiomi2014}
Y. Shiomi and E. Saitoh, 
Phys. Rev. Lett \textbf{113}, 266602 (2014).

\bibitem{MAzuma2005}
M. Azuma, K. Takata, T. Saito, S. Ishiwata, Y. Shimakawa, and M. Takano, 
J. Am. Chem. Soc., \textbf{127}, 24, 8889-8892 (2005).

\bibitem{JSu2015}
J. Su, Z. Z. Yang, X. M. Lu, J. T. Zhang, L. Gu, C. J. Lu, Q. C. Li, J.-M. Liu, J. S. Zhu, 
ACS Appl. Mater. Interfaces \textbf{7}, 24, 13260-13265 (2015).

\bibitem{WZYang}
W. Z. Yang, X. Q. Liu, H. J. Zhao, Y. Q. Lin, and X. M. Chen, 
J. Appl. Phys. \textbf{112}, 064104 (2012).

\bibitem{FGheorghiu}
F. Gheorghiu, L. Curecheriu, I. Lisiecki, P. Beaunier, S. Feraru, M. N. Palamaru, V. Musteata, N. Lupu, L. Mitoseriu, 
J. Alloys Compd \textbf{649}, 151-158 (2015).

\bibitem{SMajumder2022}
S. Majumder, M. Tripathi, R. Raghunathan, P. Rajput, S. N. Jha, D. O. de Souza, L. Olivi, S. Chowdhury, R. J. Choudhary, and D. M. Phase, 
Phys. Rev. B \textbf{105}, 024408 (2022).

\bibitem{JBGoodenough1961}
J. B. Goodenough, A. Wold, R. J. Arnott, and N. Menyuk, 
Phys. Rev. \textbf{124}, 373 (1961).

\bibitem{GBlassel1965}
G. Blasse, 
J. Phys. Chem. Solids \textbf{26}, 1969 (1965).

\bibitem{SKumar2010}
S. Kumar, G. Giovannetti, J. van den Brink, and S. Picozzi, 
Phys. Rev. B \textbf{82}, 134429 (2010).

\bibitem{WYi2013}
W. Yi, Q. Liang, Y. Matsushita, M. Tanaka, and A. A. Belik, 
Inorg. Chem. \textbf{52}, 14108 (2013).

\bibitem{SanchezBenitez2011}
J. S\'{a}nchez-Ben\'{i}tez, M. J. Mart\'{i}nez-Lope, J. A. Alonso, and J L Garc\'{i}a-Mu$\tilde{\mbox{n}}$oz, 
J. Phys.: Condens. Matter \textbf{23} 226001 (2011).

\bibitem{HNhalil2015}
H. Nhalil, H. S. Nair, C. M. N. Kumar, A. M. Strydom, and S. Elizabet, 
Phys. Rev. B \textbf{92} 214426 (2015).

\bibitem{MRetuerto2015}
M. Retuerto, \'{A}. Mu$\tilde{\mbox{n}}$oz, M. J. Mart\'{i}nez-Lope, J. A. Alonso, F. J. Mompe\'{e}n, M. T. Fern\'{a}ndez-D\'{i}az, and J. S\'{a}nchez-Ben\'{i}tez, 
Inorg. Chem. \textbf{54}, 10890-10900 (2015).

\bibitem{NTerada2015}
N. Terada, D. D. Khalyavin, P. Manuel, W. Yi, H. S. Suzuki, N. Tsujii, Y. Imanaka, and A. A. Belik, 
Phys. Rev. B \textbf{91}, 104413(2015).

\bibitem{HAdachi1999}
H. Adachi and H. Ino, 
Nature \textbf{401}, 148-150 (1999).

\bibitem{KHJBuschow1974}
K. H. J. Buschow, A. M. van Diepen, H. W. de Wijn, 
Solid St. Commun., \textbf{15}, 5 903-906 (1974) and references therein.

\bibitem{RJBooth2009}
R. J. Booth, R. Fillman, H. Whitaker, A. Nag, R. M. Tiwari, K. V. Ramanujachary, J. Gopalakrishnan, S. E. Lofland, 
Mater. Res. Bull \textbf{44}, 1559-1564 (2009).

\bibitem{PNLekshmi2013}
P. N. Lekshmi, G. R. Raji, M. Vasundhara, M. R. Varma, S. S. Pillaib and M. Valant, 
J. Mater. Chem. C, \textbf{1}, 6565-6574 (2013).

\bibitem{JLynn1990}
J. Lynn and P. Seeger, 
At. Data. Nucl. Data Tables \textbf{44}, 191 (1990).

\bibitem{VFSears1992}
V. F. Sears, 
Neutron News \textbf{3}, 26 (1992).

\bibitem{PDeBievre1993}
P. De Bievre and P. Taylor, 
Int. J. Mass Spectrom. Ion Processes \textbf{123}, 149 (1993).

\bibitem{HFischer2002}
H. E. Fischer, G. J. Cuello, P. Palleau, D. Feltin, A. C. Barnes,
Y. S. Badyal, and J. M. Simonson, 
Appl. Phys. A: Mater. Sci. Process. \textbf{74}, s160-s162 (2002).

\bibitem{MTripathi2019illdata}
M. Tripathi, T. Chatterji, H. E. Fischer, K. Schmalzl, and S. Majumder, 
Institut Laue-Langevin, https://doi.org/10.5291/ILL-DATA 5-32-869 (2019).

\bibitem{MAHowe1996}
M. A. Howe, R. L. McGreevy and P. Zetterstrom, \textit{CORRECT: a correction programme for neutron diffraction data}, 
NFL Studsvik internal report (1996).

\bibitem{JRodriguezCarvajal1993}
J. Rodríguez-Carvajal, 
Phys. B (Amsterdam, Neth.) \textbf{192}, 55 (1993).

\bibitem{JAMPaddison2013}
J. A. M. Paddison, J. R. Stewart, and A. L. Goodwin, 
J. Phys.: Condens. Matter \textbf{25}, 454220 (2013).

\bibitem{BRavel2005}
B. Ravel and M. Newville, 
J. Synchrotron Radiat. \textbf{12}, 537 (2005).

\bibitem{SMajumder2019}
S. Majumder, P. Basera, M. Tripathi, R. J. Choudhary, S. Bhattacharya, K. Bapna and D. M. Phase, 
J. Phys.: Condens. Matter \textbf{31}, 205001 (2019).

\bibitem{VMGoldschmidt1926}
V. M. Goldschmidt, 
Naturwissenschaften \textbf{14}, 477 (1926).

\bibitem{FBridges2001}
F. Bridges, C. H. Booth, M. Anderson, G. H. Kwei, J. J. Neumeier, J. Snyder, J. Mitchell, J. S. Gardner, and E. Brosha, 
Phys. Rev. B \textbf{63} (2001) 214405.

\bibitem{AHdeVries2003}
A. H. de Vries, L. Hozoi, R. Broer, 
International Journal of Quantum Chemistry \textbf{91} 57-61 (2003).

\bibitem{RIDass2003}
R. I. Dass, J.-Q. Yan, and J. B. Goodenough, 
Phys. Rev. B \textbf{68}, 064415 (2003).

\bibitem{Supplementalmaterial}
See Supplemental Material for supporting results related to magnetometric and neutron powder diffraction measurements, long range crystallographic and magnetic refinements, and real space pair distribution function calculations at different temperature values.

\bibitem{TKimura2003}
T. Kimura, S. Ishihara, H. Shintani, T. Arima, K. T. Takahashi, K. Ishizaka, and Y. Tokura, 
Phys. Rev. B \textbf{68}, 060403(R) (2003).

\bibitem{G R Haripriya2011}
G. R. Haripriya , H. S. Nair, R. Pradheesh, S. Rayaprol, V. Siruguri, D. Singh, R. Venkatesh, V. Ganesan, K. Sethupathi
and V. Sankaranarayanan, 
J. Phys.: Condens. Matter \textbf{29} 475804 (2017).

\bibitem{EBertaut1962}
E. Bertaut, 
J. Appl. Phys. \textbf{33}, 1138 (1962).

\bibitem{EBertaut1968}
E. Bertaut, 
Acta Crystallogr. A \textbf{24}, 217 (1968).

\bibitem{TEgami1991}
T. Egami, H. D. Rosenfeld, B. H. Toby, and A. Bhalla, 
Ferroelectrics \textbf{120}, 11 (1991).

\bibitem{MPasciak2019}
M. Pa\'{s}ciak et al., P. Ondrejkovic, J. Kulda, P. Van\v{e}k, J. Drahokoupil, G. Steciuk, L. Palatinus, T. R. Welberry, H. E. Fischer, J. Hlinka, and E. Buixaderas, 
Phys. Rev. B \textbf{99} 104102 (2019).

\bibitem{TEgami2012}
T. Egami, S. J. L. Billinge, 
Elsevier, ISBN: 978-0-08-097133-9 (2012).

\bibitem{DHou2018}
D. Hou, C. Zhao, A. R. Paterson, S. Li, J. L. Jones, 
J. Eur. Ceram. Soc., \textbf{38} 4, 971-987 (2018).

\bibitem{MLMedarde1997}
M. L. Medarde, 
J. Phys.: Condens. Matter \textbf{9} 1679-1707 (1997).

\bibitem{Goodenoughl1995}
J. B. Goodenough, 
Phys. Rev. \textbf{100}, 564 (1955). 

\bibitem{Kanamoril1959}
J. Kanamori, 
J. Phys. Chem. Solids \textbf{10}, 87 (1959).

\bibitem{FFan2016}
F. Fan, Z. Li, Z. Zhao, K. Yang, and H. Wu, 
Phys. Rev. B \textbf{94}, 214401 (2016).

\bibitem{RPradheesh2012}
R. Pradheesh, H. S. Nair, C. M. N. Kumar, J Lamsal, R. Nirmala1, P. N. Santhosh, W. B. Yelon, S. K. Malik, V. Sankaranarayanan, and K. Sethupathi, 
J. Appl. Phys. \textbf{111}, 053905 (2012).

\bibitem{MTripathi2017}
M. Tripathi, R. J. Choudhary, D. M. Phase, T. Chatterji, and H. E. Fischer, 
Phys. Rev. B \textbf{96}, 174421 (2017).

\bibitem{RPMadhogaria2020}
R. P. Madhogaria, E. M. Clements, V. Kalappattil, M. H. Phan, H. Srikanth, R. Das, N.T. Dang, D. P. Kozlenko, N.S. Bingham, 
J. Magn. Magn. Mater. \textbf{507}, 166821 (2020).

\bibitem{MTripathi2019}
M. Tripathi, T. Chatterji, H. E. Fischer, R. Raghunathan, S. Majumder, R. J. Choudhary, and D. M. Phase, 
Phys. Rev. B \textbf{99}, 014422 (2019).

\bibitem{SPal2019}
S. Pal, S. Jana, S. Govinda, B. Pal, S. Mukherjee, S. Keshavarz, D. Thonig, Y. Kvashnin, M. Pereiro, R. Mathieu, P. Nordblad, J. W. Freeland, O. Eriksson, O. Karis, and D. D. Sarma, 
Phys. Rev. B \textbf{100}, 045122 (2019).

\end{thebibliography}

\end{document}